\documentclass[12pt,a4paper]{article}

\usepackage[colorlinks,urlcolor=blue,linkcolor=blue,citecolor=blue]{hyperref}
\expandafter\def\expandafter\UrlBreaks\expandafter{\UrlBreaks\do\/\do\*\do\-\do\~\do\'\do\"\do\-}
\usepackage{color}
\usepackage[top=1.2in, left=1in, right=1in, bottom=1.2in]{geometry}

\usepackage{amsmath,amsfonts}
\usepackage{algorithmic}
\usepackage{array}
\usepackage[caption=false,font=normalsize,labelfont=sf,textfont=sf]{subfig}
\usepackage{textcomp}
\usepackage{stfloats}
\usepackage{url}
\usepackage{verbatim}
\usepackage{graphicx}
\def\BibTeX{{\rm B\kern-.05em{\sc i\kern-.025em b}\kern-.08emT\kern-.1667em\lower.7ex\hbox{E}\kern-.125emX}}
\usepackage{balance}
\usepackage{multirow}
\usepackage{booktabs}
\usepackage{makecell}
\usepackage{lscape}

\usepackage{tgtermes}

\newcommand\Mark[1]{\textsuperscript#1}

\AtBeginDocument{%
  \providecommand\BibTeX{{%
    Bib\TeX}}}

\begin{document}

\title{\textbf{\fontsize{20pt}{22pt}\selectfont Investigating the Perception of Facial Anonymization Techniques in 360° Videos}}
\date{}

\author{Leslie Wöhler\Mark{1}, Satoshi Ikehata\Mark{2}\Mark{,}\Mark{3}, and
Kiyoharu Aizawa\Mark{1}\\
{\small \Mark{1} The University of Tokyo, Tokyo, Japan}\\
{\small \Mark{2} National Institute of Informatics, Tokyo, Japan}\\
{\small \Mark{3} Tokyo Institute of technology, Tokyo, Japan}}

\maketitle

%%%%%%%%%%%%%%%%%%%%%%%%%%%%%%%%%%%%%%%%%%%%%%%%%%%%%%%%%%%%%%%%%%%
%%%%%%%%%%%%%%%%%%%%%%%     CONTENT   %%%%%%%%%%%%%%%%%%%%%%%%%%%%%
\begin{abstract}
In this work, we investigate facial anonymization techniques in 360° videos and assess their influence on the perceived realism, anonymization effect, and presence of participants. In comparison to traditional footage, 360° videos can convey engaging, immersive experiences that accurately represent the atmosphere of real-world locations. As the entire environment is captured simultaneously, it is necessary to anonymize the faces of bystanders in recordings of public spaces. Since this alters the video content, the perceived realism and immersion could be reduced. To understand these effects, we compare non-anonymized and anonymized 360° videos using blurring, black boxes, and face-swapping shown either on a regular screen or in a head-mounted display (HMD). 

Our results indicate significant differences in the perception of the anonymization techniques. We find that face-swapping is most realistic and least disruptive, however, participants raised concerns regarding the effectiveness of the anonymization. Furthermore, we observe that presence is affected by facial anonymization in HMD condition. Overall, the results underscore the need for facial anonymization techniques that balance both photo-realism and a sense of privacy.

\end{abstract}

\section{Introduction}
\begin{figure}
  \includegraphics[width=\textwidth]{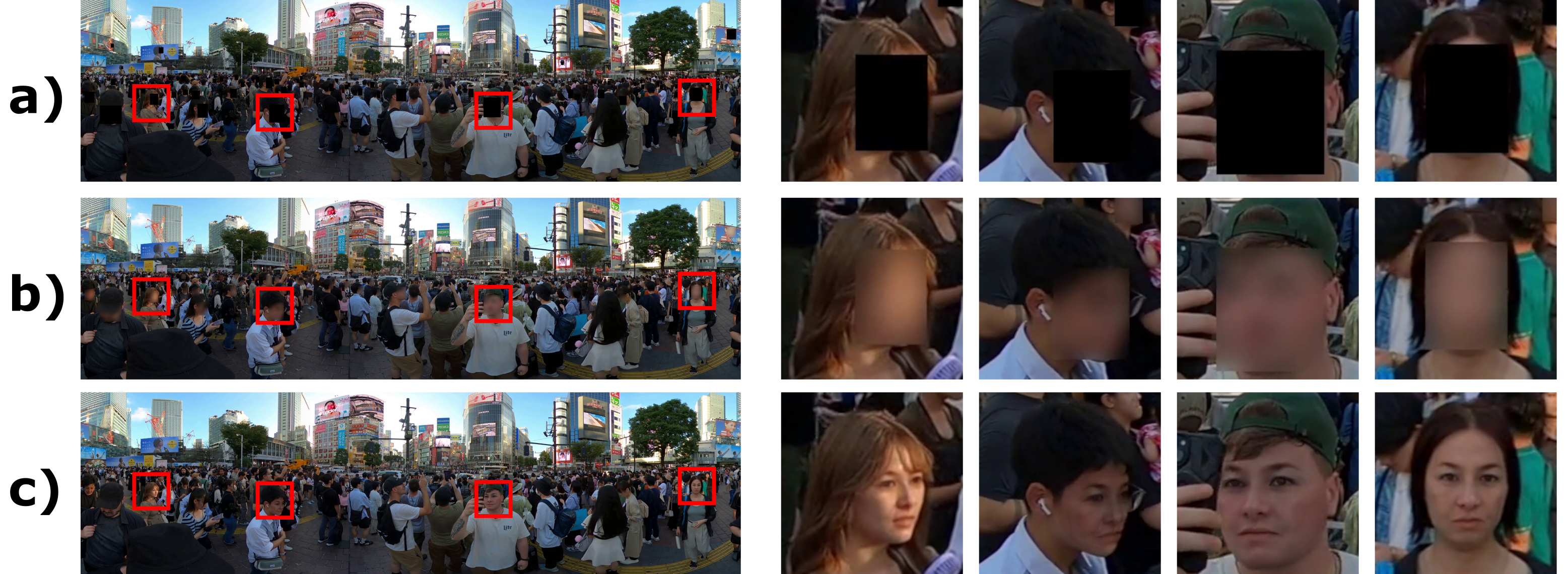}
  \caption{In this paper, we investigate three different facial anonymization techniques for 360° videos: a) Blocking of faces with black boxes, b) Blurring the facial area, c) Face-Swapping using a synthetically generated face.}
  \label{fig:teaser}
\end{figure}
Recently, the creation of 360° content has become easily feasible for non-professional users, opening countless new possibilities as they can evoke a high sense of realism, presence, and engagement~\cite{CINNAMON2023102044, narciso2019immersive, vskola2020virtual}.
As they represent the entire surrounding and allow the interactive change of viewing direction in real world locations, 360° videos are especially useful in tourism and education as they can be used to attract visitors~\cite{zhu2023exploring,rahimizhian2020emerging}, teach about remote locations~\cite{hwang:2022,ranieri2022if}, or to prepare for field trips and travel by familiarizing oneself with the environment~\cite{kumar2022tourgether360,ruberto2023comparison}.
Unfortunately, it is nearly impossible to avoid the filming of bystanders when recording 360° videos at public locations as the whole field of view is captured simultaneously.
As these bystanders have not consented to being filmed, it is mandatory to anonymize their appearance before distributing the video data.
While different techniques for the anonymization of facial features exist, they alter the video content and could impact the immersion and perceived realism of 360° videos.
Therefore, we set out to investigate the perception of facial anonymization in 360° videos.

Towards this goal, we compare current anonymization techniques to understand perceptual differences between genuine and anonymized videos.
We first record 360° videos in public spaces with varying degrees of pedestrians including shopping districts, sightseeing spots, and parks.
Afterwards, we apply three different types of facial anonymization techniques: Blocking and blurring of faces, as well as face-swapping, see Fig.~\ref{fig:teaser}.
In the past, these techniques have been explored for photo and video anonymization suggesting that blocking is especially efficient at anonymizing while blurring is more pleasant for users~\cite{vishwamitra2017blur,li2017effectiveness, fernandes2023cartoonized, mujirishvili2024don}.
Furthermore, face-swapping seems to be effective at anonymizing images~\cite{khamis:2022} and videos~\cite{wilson:2022} as it can create highly realistic results which are nearly indistinguishable from genuine recordings and preserve facial expressions~\cite{wohler:2020}.
As the effects of facial anonymization have not yet been studied for 360° videos, it is unclear how this format and immersive properties influence their perception.
Facial anonymization might become distracting when changing the view to explore the video content or lower the perceived realism and presence of participants.

In our experiment, we display the videos either on a regular screen or in a head-mounted display (HMD) as both are regularly used to view 360° videos~\cite{tse2017there,ranieri2022if}.
Participants are asked to identify the applied anonymization technique and report their impression of the anonymization and video content along with their presence based on the IPQ~\cite{ipq} questionnaire assessing spatial presence, involvement, and perceived realism.
Finally, participants directly compare the anonymization techniques regarding their distraction, effectiveness, and pleasantness after finishing all trials and express their opinions in a debriefing.
We discuss the following research questions:\begin{itemize}
    \item RQ1: Which anonymization technique is least noticeable, distracting, and most realistic?
    \item RQ2: Does facial anonymization impact the presence of participants?
    \item RQ3: Which anonymization technique is best suited for 360° videos?
\end{itemize}

To the best of our knowledge, we are the first to investigate the perception of facial anonymization in 360° videos.
Therefore, our work contributes important insights on the handling of 360° videos and design of immersive experiences.
Our results can aid creators to choose the best anonymization technique for their content and opens up discussion on how anonymization can affect the perception of users.
\section{Related Work}
In this section, we review related research on 360° videos as well as anonymization techniques for images and regular videos.

\subsection{360° Videos}
The advantages of 360° videos have been studied for various communicative and educational settings demonstrating better collaborative decision making as well as improved learning outcomes, higher engagement, and even increased empathy~\cite{ranieri2022if,hwang:2022, rosendahl2023360}. 
As they allow an interactive and realistic representation of real-world locations, they have been employed for place-based research~\cite{CINNAMON2023102044}, education~\cite{ruberto2023comparison}, and to increase awareness of environmental factors~\cite{filter2020virtual}.
Their benefits include their ease of use and low cost, as well as interactivity, and the possibility to convey a sense of immersive realism and presence~\cite{narciso2019immersive, vskola2020virtual}.
Furthermore, 360° videos of real world locations can be enriched by introducing further interaction and collaboration possibilities to create virtual tours or photo-realistic virtual environments~\cite{takenawa2023360rvw, sugimoto2020building, kayukawa20233dmoviemap, kumar2022tourgether360}.
This way, they enhance the engagement and presence of users to bridge the gap between videos and 3D generated virtual environments.

There has been extensive research on the presence of users in virtual environments which found many influencing factors based on the display~\cite{teixeira2021effects}, interactivity~\cite{sanchez2005presence}, and realism~\cite{slater2009visual, Crescent}.
In general 360° videos in HMDs evoke highest immersion~\cite{tse2017there}, but require specialized hardware and might lead to discomfort or cybersickness~\cite{ranieri2022if}.
Therefore, it can be beneficial to view 360° videos on a regular screen as this can still offer increased engagement~\cite{rahimizhian2020emerging} and a satisfying experience with a feeling of presence~\cite{zhu2023exploring}.

As 360° videos can offer rich experiences when viewed on either regular screens or in HMDs, we aim to investigate both conditions.

\subsection{Facial Anonymization}
In general, privacy concerns regarding images and videos recorded in public spaces have been discussed for various scenarios including surveillance systems~\cite{nguyencctv,newton2005preserving}, drone recordings~\cite{wang2016flying}, and live streaming~\cite{faklaris2020snapshot} which contributed to the development of automatized anonymization systems for large scale image collections of public locations~\cite{frome2009large,nodari2012digital, gallo2020privacy}.
Oftentimes, bystanders feel unwell about being recorded and worry about the usage of their video data~\cite{faklaris2020snapshot}.
While visible cameras lead to decreased concerns as people can avoid them~\cite{singhal2016you}, this is not easily possible for 360° videos.

To circumvent privacy concerns, different anonymization techniques have been introduced and evaluated for images and videos.
One study showed either original or anonymized images (blurring or blocking) asking participants to identify the anonymized person, rate their satisfaction with the anonymization, how much they liked the image, the information sufficiency, and their sense of social presence~\cite{vishwamitra2017blur}.
It was found that blocking image regions is most effective, however, blurring was rated more positive on all other variables.
Another work employed a wide array of anonymization techniques and directly asked participants about their opinion on the utilized anonymization technique regarding its likability and their preference revealing that completely removing people using in-painting or cartoon avatars offers a pleasant viewing experience and ensures privacy~\cite{li2017effectiveness}.

The perception of facial anonymization of videos has been studied for various contexts often discussing the trade-off between video context and privacy preservation~\cite{boyle2005language}.
For interviews, anonymization using AI-based stylization was proposed as a method to ensure privacy while allowing to convey emotional information~\cite{yalccin2024empathy}.
In the study, participants watched videos anonymized using stylization as well as common techniques and answered questions on perspective taking and empathetic concern.
The results indicate similar perception between the stylization and common anonymization techniques, however, they differ from unaltered videos.
For streaming, cartoon-based anonymization that offers strong anonymization while preserving the video context was proposed, however, viewers still favored the non-anonymized videos~\cite{hasan2017cartooning}.
In contrast, work on ego-centric cameras showed that both wearers of the cameras value privacy preservation and preferred cartooning over blurring and blocking~\cite{fernandes2023cartoonized} as it keeps the video context.
For assisted-living videos, patients were testing various anonymization techniques including blurring or complete removal and healthcare providers watched the resulting videos~\cite{mujirishvili2024don}.
The authors performed semi-structured interviews with the participants reveling that the preferred privacy protection technique can not only vary between stakeholders but also based on the location of the camera (e.g., bathroom vs. living room).

Due to its natural appearance, unobtrusiveness, and ability to preserve facial expressions~\cite{wohler:2021, woehler:2022, tahirdeepfake}, face-swapping has been explored as an anonymization technique.
It has been studied on videos of children used to research and diagnose autism~\cite{wilson:2022}. 
The authors computationally evaluated the accuracy of the face-swaps gaze and anonymization efficiency concluding that despite some inaccuracies, it would be a valuable trade-off for videos that require strict privacy preservation.
In another study, participants were not informed of the anonymization and asked to identify face-swapped public figures.
The results show that face-swapping leads to high anonymization success~\cite{khamis:2022}.

To the best of our knowledge, facial anonymization techniques have not yet been assessed for 360° videos.
\section{Method}
To assess our research questions, we first formulate hypotheses and afterwards design our stimuli and experiment.

\paragraph*{Hypotheses.}
Due to the immersive nature of 360° videos, we assume that conspicuous facial anonymization techniques can lead to reduced realism and changes in perception.
To this end, we compare the three anonymization techniques blocking, blurring, and face-swapping (see Fig.~\ref{fig:teaser}).
As face-swapping has been found to be non-obtrusive in portrait videos~\cite{wohler:2021}, we estimate that it is also difficult to notice in 360° videos.
Due to this, face-swapping might be perceived similar to the original videos.
In contrast, blocking the faces with black rectangles might be easily visible and therefore might be perceived as the most effective anonymization~\cite{li2017effectiveness}.

Overall, we aim to evaluate the following hypotheses:
\begin{itemize}
    \item H1: Face-swapping is difficult to identify and best at preserving the realism of the videos.
    \item H2: Blocking is perceived as most effective at anonymizing the faces.
    \item H3: Face-swapping is least distracting, blocking is most distracting.
    \item H4: Face-swapping is most pleasant and preferred by participants.
    \item H5: Facial anonymization techniques influence presence and the impression the scene has on viewers.
\end{itemize}
\begin{figure}
    \centering
\includegraphics[width=\textwidth]{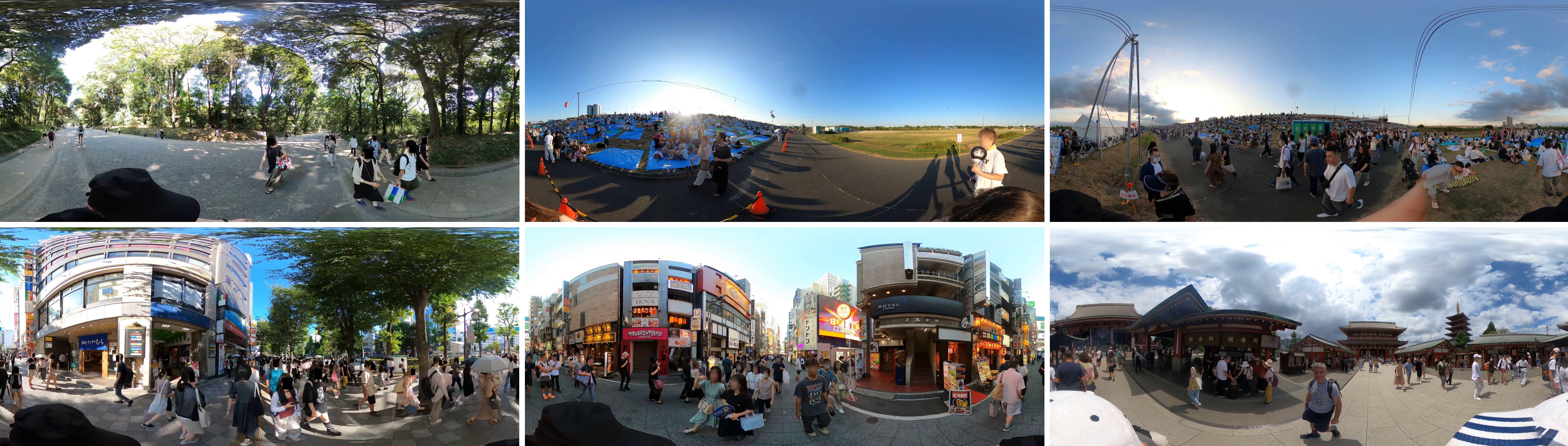}
    \caption{Exemplar stimuli of different scenes with varying amounts of bystanders (Left: Block, Middle: Blur, Right: Swap).}
    \label{fig:stimuli_example}
\end{figure}

\begin{figure*}
    \centering
    \includegraphics[width=\textwidth]{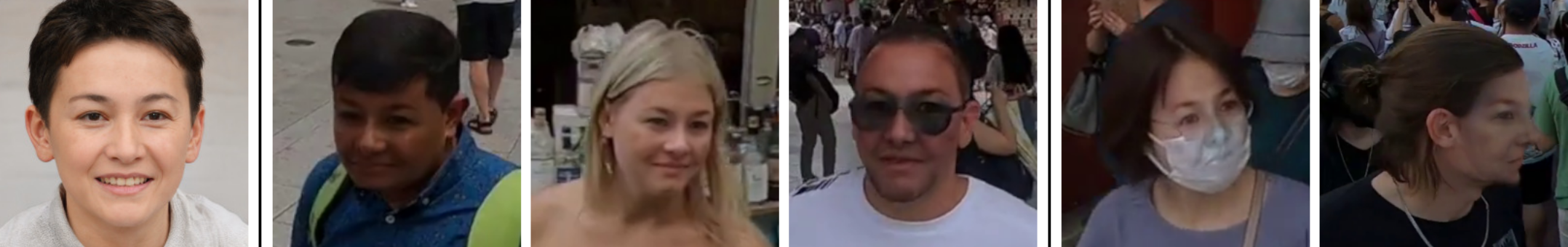}
    \caption{A synthetic face (left) is used to swap all faces in the 360° videos. While this produces high-quality results for different people and minor occlusions (middle) artifacts can occur if large areas of the face are occluded and in profile views (right).
    }
    \label{fig:syntheticface}
\end{figure*}

\paragraph*{Stimuli.}
We create our stimuli by first recording a total of 16 scenes in 4k resolution and 360° format at public locations at shopping districts, sightseeing spots, and parks using a GoPro Max 360° camera.
The videos show varying amount of pedestrians from highly crowded areas to scenes with only a few faces visible at the same time as shown in Fig.~\ref{fig:stimuli_example}.
Afterwards, we cut the video of each scene to a length of 30 seconds and prepare stimuli variations by anonymizing the facial identities in the videos (Fig.~\ref{fig:teaser}).
To create the anonymizations, we first perform face detection using RetinaFace~\cite{retinaface}.
We manually choose the face detection threshold for each video in a way that most faces are detected.
For blurring, Gaussian blur with kernel size of 71 pixel is applied in the detect area, for blocking the area is colored black.
For face-swapping the SimSwap framework~\cite{simswap} is used as it allows to replace arbitrary faces without training for specific identities while keeping gaze and facial expressions intact. 
We replace all detected faces with the same synthetically generated appearance which can retain some of the original persons characteristics like their skin color or glasses while preserving the target identity (see Fig.~\ref{fig:syntheticface}).

This way, we acquire a total of 64 stimuli (16 videos each in condition original, blocking, blurring, face-swapping).
In the experiment, we mute the audio so participants focus only on the visual information.

\paragraph{Apparatus.}
The experiment is conducted in two different settings: Using a regular screen, or HMD.
Experiments for the Screen condition are performed using Amazon Mechanical Turk, while the HMD condition is conducted in-situ using a PICO 4 Enterprise.
In the Screen condition, participants can change the viewing direction in the video by dragging  the mouse, in the HMD condition participants are seated on a rotating office chair so they can easily change the viewing direction by rotating their head or whole body.
The controls are explained at the beginning of the experiment.
In the Screen condition, we instruct participants to use a sufficient screen (i.e., computer monitor) and high-speed internet connection in order to correctly display the stimuli,  however, we cannot control the actual screen size and streaming quality.

\paragraph{Measures.}
We aim to measure the opinion of participants on the different anonymization techniques as well as their sense of presence and impression of the scene.

After each trial, we ask whether the video was anonymized and which anonymization technique they think was used which can inform us on how easy it is to identify each technique.
For each trial, we also ask whether the anonymization in the video was efficient, distracting, and whether it reduced the realism of the video (7-point Likert scale).
Next to the direct assessment of the anonymization, we also aim to evaluate the presence of participants. According to previous work, presence itself results from immersion and can be described by various behaviors and impressions~\cite{slater2003note}.
We use the IPQ questionnaire~\cite{ipq, ipq2, ipq3} which specifically assesses spatial presence (i.e., feeling physically present in the scene), involvement (i.e., attention to the virtual environment and experience of involvement), as well as the experienced realism (i.e., the subjective impression of the scenes realism).
The IPQ questionnaire was throughout evaluated and found to have high reliability~\cite{schwind}.
Next to presence, we also aim to understand whether the impression and emotions of participants are affected by anonymization as applications of 360° videos might seek to elicit specific responses, e.g., portraying a relaxing and attractive tourist destination, or an interesting and comfortable experience for educational purposes.
Specifically, we ask about the attractiveness of the location, the atmosphere, the crowding, as well as four adjective pairs (Stressful/Relaxable, Boring/ Interesting, Uncomfortable/ Comfortable, Unfamiliar/ Familiar) each rated on a 7-point Likert scale.

We additionally include a post experiment questionnaire which asks participants to directly rate the three different anonymization techniques.
Specifically, participants select which technique they felt was overall most/ least distracting, effective, pleasant, and which technique they generally prefer.
Finally, we ask whether they feel that everyone had the same face in face-swapped videos to measure the impact of only using one synthetic facial appearance for all face-swaps.
In the HMD condition, we additionally conduct a debriefing about general impressions of the anonymization techniques following the format of a structured interview.
We first ask participants to describe their general impression of blocking, blurring, and face-swapping.
Afterwards, we ask them which anonymization technique should be used for 360° videos and to provide a reasoning for their decision.
The experiment conductor transcribes the answers of participants.
This format allows us to gain more insights into the perception of the anonymization techniques as participants can freely state their observations.

The experiment in screen condition was conducted in English, while the in-situ experiment uses a Japanese translation.
For the presence questionnaire~\cite{ipq, ipq2, ipq3}, the English and Japanese translation proposed on the IPQ website were used.
All other questions were translated from English to Japanese in cooperation with a native Japanese speaker.
The debriefing was conducted and transcribed in Japanese - the native language of participants and experiment conductor - and afterwards translated to English for the reports in this paper.
The experiment was permitted by the ethical committee of our university.

\paragraph*{Participants.}
We recruit 20 participants for both the Screen and HMD condition.
There is no overlap between participants in both conditions.
For the Screen condition, we recruit participants from the US using Amazon Mechanical Turk.
Participants (13 female, 7 male) have an average age of 31.65 years (SD 8.99) with ages between 22 and 50.
The experiment takes around 20 minutes and they receive 5 USD as compensation.
For the HMD condition, we recruit participants (9 female, 10 male, 1 prefer not to say) from our university.
They have an average age of 23.35 years (SD 1.5) with ages between 21 and 26.
The experiment takes around 50 minutes and they receive a gift card with a value of 1500 Yen as compensation.

\paragraph*{Procedure.}
We use a counter-balanced between-participant design with randomization for our experiment.
Every participant watches all 16 videos, each only once in one of the four conditions (Either original, blocking, blurring, or face-swap).
Additionally, we balance the stimuli selection so that each condition is shown four times to each participant.
Overall, we also ensure to receive the same amount of annotations for each video-condition combination.

Before the experiment, we explain the goal and content of the experiment, obtain informed consent, collect demographic information and ask about previous experience with VR content.
Furthermore, we demonstrate how to change the viewing direction in 360° videos, explain the four possible anonymization conditions and show an example image for each technique.

During the trials, first a stimuli video is shown to participants without any additional information.
Participants are not able to rewind or pause the videos.
After the video has ended, it is no longer visible and instead the questionnaire is displayed.
Once the participant answered all questions, they can start the next video by pressing a button.
This procedure is repeated for the 16 trials.
Once all trials are completed, the post experiment questionnaire is displayed.
Additionally, the participants are asked whether they have previously visited any of the locations shown in the videos.
In the HMD condition, the experiment ends with the debriefing.
\section{Analysis and Results}
Before the analysis of our hypotheses, we verified that our data satisfies the assumptions of normality and the homogeneity of variances (Shapiro-Wilk, Levene).

\paragraph{Correct Identification of Anonymization Techniques.}
After each video, participants were asked to report whether the video has been anonymized and if so, which technique has been used.
In both conditions, the identification of blocking was easiest with correct identification in 91.25\% of trials in Screen and 96.25\% in HMD condition, followed by blurring (Screen 86.25\%, HMD 85.0\%), and face-swapping (Screen 32.5\%, HMD 42.5\%).
In HMD condition, original videos were reported correctly in 87.5\% of the videos.
In Screen condition, participants sometimes confused them with face-swapped and blurred videos leading to a lower accuracy of 58.7\%. 
Furthermore, participants did not easily recognize that only one facial identity was used rating the statement "In face-swapped videos, everyone had the same face." with mean ratings of 0.05 in Screen (SEM 0.42) and -1.55 in HMD (SEM 0.32) ( -3 = Fully Disagree, 3 = Fully Agree).

These results indicate that despite the additional challenges of 360° videos like profile views and varying lighting conditions, face-swaps are still difficult to recognize \textbf{supporting hypotheses H1 that face-swapping is difficult to identify}.
\begin{figure*}[t]
    \centering
    \includegraphics[width=\textwidth]{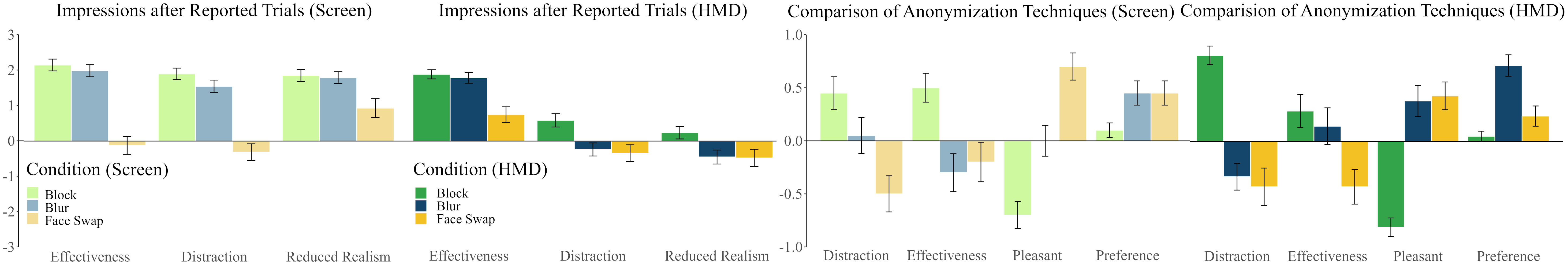}
    \caption{Impression of the anonymization techniques in HMD and Screen condition. Left: Average ratings for trials correctly reported to be anonymized. Right: Answers for the comparison of techniques in the post experiment questionnaire. Error bars indicate the standard error of the mean (SEM).}
    \label{fig:combined}
\end{figure*}

\paragraph{Perception of the Anonymization Techniques.}
After each trial, we asked participants about their impression of the anonymization in that specific video regarding the effectiveness and distraction of the anonymization as well as whether they think the realism was reduced in the videos.
To assess the impression of the anonymization techniques, we first evaluate only trials that were correctly identified as anonymized.
Overall, we find that in both Screen and HMD condition, blocking is rated as most effective, distracting, and least realistic followed by blurring, and face-swapping, see Fig.~\ref{fig:combined} (left).

For the Screen condition, our analysis indicates significant differences between the three anonymization techniques (Tab.~\ref{Tab:tests}).
Post-hoc tests indicate that the ratings for face-swapped videos are significantly different from both other conditions (Tab.~\ref{Tab:Posthoc}).
So it seems participants in Screen condition found the anonymization of face-swapping less efficient than the other techniques, however, they also felt it was less distracting and did not impact the realism of the video as much.

For the HMD condition, we also find differences between the conditions and variables (Tab.~\ref{Tab:tests}).
However, post-hoc tests only reveal differences between face-swapping and the other two techniques regarding their effectiveness.
We also find that blocking is most distracting and less realistic than blurring (Tab.~\ref{Tab:Posthoc}).
So in general participants in HMD condition found face-swapping less effective, while rating the distraction and realism similar to blurring indicating that face-swapping is less suited for videos watched in HMD than blurring.

As face-swapping was often not recognized by participants, we also look into the differences in responses for trials in which videos were reported as anonymized and not reported as anonymized.
We find that videos were rated as less distracting, and more realistic if participants did not notice the anonymization.
Especially in HMD condition, this impacts the analysis and results in face-swapped trials to be significantly less distracting and more realistic than blurring and blocking (Tab.~\ref{Tab:Posthoc}).

From these results, we can see that \textbf{trials using face-swapping were rated as most realistic and least distracting in Screen condition supporting H1 and H3}.
When looking at all trials, we furthermore find high realism and lowered distraction for face-swapping in HMD condition.
This indicates that visible artifacts can arise during face-swapping which not only makes it possible to detect the technique but also distract participants and lowers the perceived realism.
Moreover, our results \textbf{do not support hypotheses H2 as blocking is not rated as most effective} at anonymizing the faces.

\paragraph{Comparison of the Anonymization Techniques.}
In the post experiment questionnaire, participants selected their preferred technique and ranked the distraction, effectiveness, and pleasantness as shown in Fig.~\ref{fig:combined} (right).
For the Screen condition, our statistical assessment indicates differences between all three techniques for all questions and furthermore highlights differences for each question individually (Tab.~\ref{Tab:tests}).
Post-hoc test show that blocking is more distracting than swapping, most effective, least pleasant, and least preferred, while face-swapping was perceived as most pleasant (Tab. ~\ref{Tab:Posthoc}).

For the HMD condition, we similarly find differences between the three techniques, with following test indicating differences for all variables (Tab.~\ref{Tab:tests}).
Similar to the Screen condition, post-hoc test with Tukey HSD indicate that blocking is most distracting and least pleasant, while face-swapping is rated as least efficient.
In HMD condition, participants seem to overall prefer blurring as anonymization technique.

These \textbf{results support H2 and H3 for videos shown in Screen condition as blocking received the highest ratings for effectiveness and was perceived as most distracting while face-swapping was judged to be least distracting.}
For HMD condition, we cannot detect significant differences regarding the effectiveness of blocking and blurring or the distraction of blurring and face-swapping.
Additionally, we find \textbf{only partial support for H4.
Against our hypothesis, face-swapping is rated as most pleasant in Screen condition but not in HMD condition.}
Furthermore, despite the high pleasantness it is not overall preferred by participants.
Instead, it is similarly rated to blurring in Screen condition and in HMD condition blurring is overall preferred.
\begin{figure*}
\includegraphics[width=\textwidth]{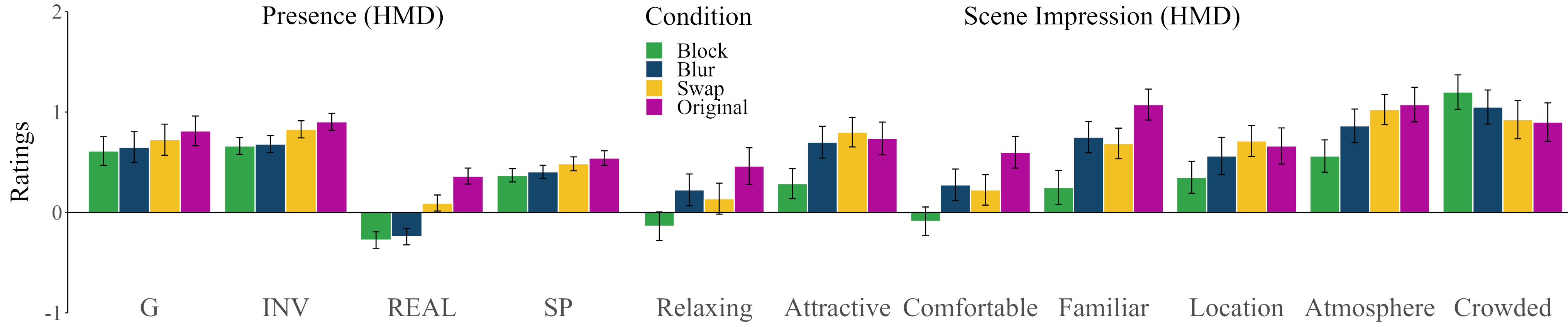}
    \caption{Mean ratings for presence (IPQ scales: G - General Presence, INV - Involvement, REAL - Experienced Realism, SP - Spatial Presence) and impressions of the scene in HMD condition. Error bars indicate the SEM.}
    \label{fig:presenceperception}
\end{figure*}

\paragraph{Presence and Scene Impression.}
Next, we asses the differences in presence and scene impressions.
For the Screen condition, our statistical analysis does not support differences between the conditions for presence and scene impressions.
For the HMD condition, we can see a general trend of higher ratings on the presence scales for the original videos, followed by face-swapping, blurring, and blocking (Fig.~\ref{fig:presenceperception}).
Our assessment points towards significant differences between the conditions on the realism scales with post-hoc test indicating that original and face-swapped videos retain a higher realism than blocked or blurred videos (Tab.~\ref{Tab:tests}, Tab. ~\ref{Tab:Posthoc}).

For participants impression of the scene, we find differences for the rating of adjective pairs comfortable/ uncomfortable, familiar/ unfamiliar with post-hoc test showing differences between blocked and non-anonymized videos.
Thereby, original videos are rated more positive than blocked videos indicating a negative influence of blocking on the scene impression.

Overall this means \textbf{we only find support for hypothesis H5 that anonymization techniques influence presence and scene impression for videos watched in HMD}.
Additionally, it seems as if blocking reduces positive impressions of the scene, while face-swapping might be best at preserving the presence of users.

\begin{table}
\resizebox{0.95\textwidth}{!}{%
\begin{tabular}{|l|l|l|l|l|l|l|l|l|}
\hline
\textbf{Question}           & \textbf{Variable}               & \textbf{Data} & \textbf{Cond.} & \textbf{Test}& \textbf{F} & \textbf{dfn} & \textbf{dfd} & \textbf{p} \\
\hline 

Perception of Anonymization & All &RT& Screen &MANOVA& 3.93 & 2 & 376 & $<$ 0.001\\
Perception of Anonymization &Effective &RT &Screen &ANOVA & 9.34 & 2 & 189 & $<$ 0.0002\\
Perception of Anonymization &Distraction &RT &Screen &ANOVA & 8.53 & 2 & 189 & $<$ 0.0003\\
Perception of Anonymization&Realism &RT &Screen &ANOVA& 5.75 & 2 & 189 & $<$ 0.004\\

Perception of Anonymization & All &AT& Screen &MANOVA& 14.02 & 2 & 472 & $<$ 1.0e-6\\
Perception of Anonymization &Effective &AT &Screen &ANOVA & 40.4 & 2 & 237 & $<$ 1.0e-5\\
Perception of Anonymization &Distraction &AT &Screen &ANOVA &37.76 & 2 & 237  & $<$ 0.0001\\
Perception of Anonymization&Realism &AT &Screen &ANOVA& 32.47 & 2 & 237  & 1.0e-5\\
\hline

Perception of Anonymization&All &RT &HMD& MANOVA& 5.93 & 2 & 400 & $<$ 0.0001\\
Perception of Anonymization&Effective &RT &HMD &ANOVA& 11.98 & 2 & 201 & $<$ 0.0001\\
Perception of Anonymization& Distraction &RT &HMD &ANOVA& 6.72 & 2 & 201 & $<$ 0.001\\
Perception of Anonymization&Realism &RT &HMD &ANOVA& 4.27 & 2 & 201 & $<$ 0.02\\
\hline

Perception of Anonymization&All &AT &HMD& MANOVA& 16.03 & 2 & 472 & $<$ 0.0001\\
Perception of Anonymization&Effective &AT &HMD &ANOVA& 39.6 & 2 & 237 & $<$ 1.0e-5\\
Perception of Anonymization& Distraction &AT &HMD &ANOVA& 24.18 & 2 & 237 & $<$ 0.001\\
Perception of Anonymization&Realism &AT &HMD &ANOVA& 16.25 & 2 & 237 & $<$ 1.0e-5\\
\hline

Comparison of Anonymization  &All &Post. &Screen &MANOVA& 7.75 & 2 & 110 & $<$ 1.0e-6\\
Comparison of Anonymization&Effective &Post. &Screen &ANOVA& 6.67 & 2 & 57 & $<$ 0.002\\
Comparison of Anonymization& Distraction&Post. &Screen&ANOVA&  8.39 & 2 & 57 & $<$ 0.0001\\
Comparison of Anonymization&Pleasant &Post. &Screen&ANOVA & 27.38 & 2 & 57 & $<$ 1.0e-6\\
Comparison of Anonymization&Preference &Post. &Screen &ANOVA&3.98 & 2 & 57 & $<$ 0.03\\

Comparison of Anonymization& All &Post. &HMD &MANOVA& 13.01 & 2 & 116 & $<$ 1.0e-6\\
Comparison of Anonymization&Effective &Post. &HMD &ANOVA& 5.29 & 2 & 60 & $<$ 0.008\\
Comparison of Anonymization& Distraction &Post. &HMD &ANOVA& 25.95 & 2 & 60 & $<$ 1.0e-5\\
Comparison of Anonymization&Pleasant &Post. &HMD &ANOVA& 32.07 & 2 & 60 & $<$ 1.0e-5\\
Comparison of Anonymization&Preference &Post.& HMD&ANOVA&  16.42 & 2 & 60 & $<$ 0.0001\\
\hline

IPQ &All &RT &HMD&MANOVA&  1.8 & 3 & 1276 & $<$ 0.04\\
IPQ &Realism &RT &HMD &ANOVA& 13.87 & 3 & 945 & $<$ 1.0e-6\\
Scene Impression &Familiar &RT &HMD &ANOVA & 3.4 & 3 & 316 & $<$ 0.02\\
Scene Impression &Comfortable &RT &HMD &ANOVA& 4.63 & 3 & 316 & $<$ 0.005\\
\hline

\end{tabular}}
\caption{Significant results of the performed MANOVA and ANOVA tests (RT~=~Reported trial, AT~=~All trials, Post.~=~Post experiment questionnaire).}
\label{Tab:tests}
\end{table}

\begin{table}
\resizebox{0.95\textwidth}{!}{%
\begin{tabular}{|l|l|l|l|l|l|l|}
\hline
\textbf{Question}      & \textbf{Variable}     & \textbf{Data}               & \textbf{Cond.} & \textbf{Anon.} & \textbf{p} & \textbf{CI} [Lo, Hi] \\
\hline 
Perception of Anonymization &    Effective & Reported Trials       &  Screen  & FS, BB & 0.0003 &  [-1.9, -0.49]  \\ 
 Perception of Anonymization &   Effective    & Reported Trials     &  Screen  & FS, B  & 0.0006 &  [-1.84, -0.42] \\ 
  Perception of Anonymization &  Distraction    & Reported Trials    &  Screen  & FS, BB & 0.0002 &  [-1.88, -0.5] \\ 
    Perception of Anonymization &Distraction   & Reported Trials     &  Screen  & FS, B  & 0.006 &  [-1.61, -0.22] \\
    Perception of Anonymization & Realism  & Reported Trials  &  Screen  & FS, BB & 0.005 &  [-1.61, -0.23] \\ 
    Perception of Anonymization & Realism  & Reported Trials   &  Screen  & FS, B  & 0.01 &  [-1.56, -0.17] \\   
   Perception of Anonymization &Effective  &  All Trials    &  Screen  & FS, BB & 1.7e-13 &  [-2.94, -1.61]  \\ 
    Perception of Anonymization &Effective   &  All Trials     &  Screen  & FS, B  & 4.4e-12 &  [-2.78, -1.45] \\ 
    Perception of Anonymization &Distraction   &  All Trials    &  Screen  & FS, BB & 1.6e-13 &  [-2.86, -1.57] \\ 
    Perception of Anonymization &Distraction  &  All Trials     &  Screen  & FS, B  & 2.5e-10 &  [-2.51, -1.22] \\ 
     Perception of Anonymization &Realism &  All Trials   &  Screen  & FS, BB & 3.6e-11 &  [-2.64, -1.33] \\ 
     Perception of Anonymization &Realism &  All Trials  &  Screen  & FS, B  & 2.3e-10 &  [-2.56, -1.24] \\ \hline  
   Perception of Anonymization & Effective  & Reported Trials      &  HMD  & FS, BB & 2.0e-5 &  [-1.72, -0.56]  \\ 
    Perception of Anonymization &Effective   & Reported Trials     &  HMD  & FS, B  & 1.1e-4 &  [-1.62, -0.46] \\ 
   Perception of Anonymization & Distraction   & Reported Trials    &  HMD  & FS, BB & 0.008 &  [-1.66, -0.2] \\ 
  Perception of Anonymization &  Distraction   & Reported Trials    &  HMD  & BB, B  & 0.005 &  [-1.44, -0.21] \\ 
    Perception of Anonymization & Realism  & Reported Trials &  HMD  & BB, B & 0.03 &  [-1.3, -0.07] \\  
Perception of Anonymization &    Effective  &  All Trials       &  HMD  & FS, BB & 3.3e-13 &  [-2.35, -1.27]  \\ 
   Perception of Anonymization & Effective   &  All Trials      &  HMD  & FS, B  & 4.4e-12 &  [-2.25, -1.17] \\ 
  Perception of Anonymization &  Distraction  &  All Trials      &  HMD  & FS, BB & 1.1e-10 &  [-2.43, -1.2] \\ 
  Perception of Anonymization &  Distraction  &  All Trials      &  HMD  & FS, B  & 5.7e-04 &  [-1.6, -0.37] \\ 
  Perception of Anonymization &  Distraction  &  All Trials      &  HMD  & BB, B  & 5.0e-3 &  [-1.44, -0.21] \\ 
 Perception of Anonymization &     Realism &  All Trials   &  HMD  & FS, BB & 1.1e-7 &  [-2.12, -0.87] \\ 
  Perception of Anonymization &    Realism &  All Trials   &  HMD  & FS, B  & 6.4e-3 &  [-1.4, -0.19] \\ 
Perception of Anonymization &     Realism  &  All Trials  &  HMD  & BB, B  & 2.6e-2 &  [-1.31, -0.06] \\ 
\hline 
IPQ & REAL   &  All Trials  &  HMD  & FS, BB & 7.2e-3 &  [0.07, 0.66] \\  
    IPQ& REAL  &  All Trials   &  HMD  & FS, B & 0.02 &  [0.04, 0.63] \\  
    IPQ &REAL  &  All Trials   &  HMD  & O, BB & 11.9e-7 &  [0.34, 0.93] \\  
    IPQ &REAL  &  All Trials   &  HMD  & O, B & 9.7e-7 &  [0.31, 0.9] \\  
    Scene Impression &Comfort  &  All Trials   &  HMD  & O, BB & 0.009 &  [0.13, 1.25] \\ 
    Scene Impression &Familiar   &  All Trials  &  HMD  & O, BB & 0.001 &  [0.25, 1.4] \\  
    \hline  
  Comparison of Anonymization  & Distraction & Post.  &  Screen  & BB, FS & 4.1e-4 &  [-1.51, -0.39] \\  
  Comparison of Anonymization &  Effective & Post.   &  Screen  & BB, B & 0.004 &  [-1.37, -0.22] \\ 
   Comparison of Anonymization&  Effective & Post.   &  Screen  & BB, FS & 0.02 &  [-1.27, -0.13] \\  
  Comparison of Anonymization &  Pleasant & Post.   &  Screen  & BB, B & 0.02 &  [0.24, 1.16] \\  
  Comparison of Anonymization &  Pleasant & Post.   &  Screen  & BB, FS & 2.1e-9 &  [0.94, 1.86] \\ 
  Comparison of Anonymization &  Pleasant  & Post.  &  Screen  & B, FS & 0.02 &  [0.25, 1.16] \\  
    
   Comparison of Anonymization & Distraction & Post.  &  HMD  & BB, FS & 5.9e-8 &  [-1.7, -0.78] \\  
  Comparison of Anonymization &  Distraction & Post.   &  HMD  & BB, B & 4.0e-7 &  [-1.6, -0.68] \\  
   Comparison of Anonymization&  Effective & Post.   &  HMD  & BB, FS & 0.008 &  [-1.27, -0.16] \\  
  Comparison of Anonymization &  Effective & Post.  &  HMD  & B, FS & 0.04 &  [-1.13, -0.01] \\  
 Comparison of Anonymization &   Pleasant & Post.   &  HMD  & BB, B & 1.7e-8 &  [0.77, 1.61] \\  
  Comparison of Anonymization &  Pleasant & Post.   &  HMD  & BB, FS & 5.8e-9 &  [0.81, 1.66] \\  
Comparison of Anonymization  &  Preference  & Post.  &  HMD  & B, BB & 1.9e-6 &  [0.37, 0.95]] \\  
  Comparison of Anonymization &  Preference & Post.  &  HMD  & B, FS & 5.6e-4 &  [-0.76, -0.19] \\ \hline 

\end{tabular}}
\caption{Significant results for the performed post-hoc test with Tukey HSD. (O~-~Original, BB~-~Blocking, B~-~Blurring, FS~-~Face-swapping; Post.~-~Post experiment questionnaire)}
\label{Tab:Posthoc}
\end{table}

\paragraph{Debriefing.}
During the debriefing in HMD condition, we performed a structured interview consisting of one question about the general impression for each of the anonymization techniques and one question asking which of the anonymization techniques should be used for 360° videos.

To assess the data, we first analyze the transcribed content and identify keywords participants used to describe the anonymization techniques.
We find that participants generally describe the naturalness (e.g., P03:"Blocking is lacking in realism", P09: "Blur looked natural"), distraction (e.g., P11: "Blur does’t detract from the scene", P19: "Face-swapping does not interfere with my enjoyment of the video"), discomfort (e.g., p06: "Face-swapping was unsettling", P09: "Blocking was bothersome"), anonymization effectiveness (e.g., P03: "As for face-swapping, I cannot judge whether it is suitable [as anonymization] or not because I didn’t notice it in videos.", P19: "From the perspective of anonymization, blocking is the best.") and artifacts (e.g., P06: "Face-swapping sometimes didn’t overlap well", P07: "Blocking flickers").

Specifically, we find that participants described blocking as mostly negative, referring to it as unnatural (9 reports), distracting (10 reports), as well as describing discomfort (10 reports) and artifacts (10 reports).
One participant even stated that "black bars reminded me of censorship practices and gave a negative impression due to the sense of oppression it conveyed" (P18).
In contrast, blurring is viewed as more positive regarding the naturalness (7 reports) and seen as non-distracting (8 reports).
Furthermore, three participants referred to blurring as familiar since the technique is often used on TV.
Face-swapping was described as natural 11 times, while discomfort was reported 8 times.
This may indicate that face-swapping usually becomes disruptive when unappealing artifacts are encountered as was also stated directly by 5 participants.
Moreover, one participant said face-swapping increases discomfort if the face does not match the ethnicity of the person (P4) and one participant described the occurrence of several people with the same face as scary (P11).
Regarding the effectiveness of the anonymization, participants felt it was difficult to judge face-swapping as it was hard to identify, so they wondered whether faces were actually anonymized sufficiently (6 reports).
For all techniques, flickering of faces was the most reported artifact which occurs when face detection fails (Blocking 9, Blurring 3, Face-swapping 4 reports).

In the last question, 12 participants said blurring should be used to anonymize 360° videos because of the natural impression (6), minimal distraction (5), reduced artifacts (2), and effectiveness (1).
Furthermore, 6 participants would choose face-swapping due to the highly natural appearance (4) and reduced distraction (3).
The last 2 participants said that either blurring or face-swapping should be used as they both retain the natural impression of the scene.

\paragraph{Summary.}
In summary, we evaluate our hypotheses as follows.

\begin{itemize}
    \item H1: Face-swapping is difficult to identify and best at preserving the realism of the videos:
    We find that face-swapping is the least noticed technique, received better ratings for realism, and was most often described as natural. 
   \item H2: Blocking is perceived as most effective at anonymizing the faces:
    We find no support for H3 as blocking is only rated as most effective during the direct comparison of techniques in Screen condition.
    \item H3: Face-swapping is least distracting, blocking is most distracting:
    We find partial support for this hypothesis, as face-swapping is least distracting in Screen condition and blocking is most distracting in HMD condition.
    \item H4: Face-swapping is most pleasant and preferred by participants.
    We find no support for this hypothesis. While face-swapping is most pleasant in Screen condition, it is not preferred by participants.
    \item H5: Facial anonymization techniques influence presence and the impression the scene has on viewers:
    We find partial support for this hypothesis, as the presence and scene impression is affected by anonymization in HMD condition.
\end{itemize}
\section{Discussion}
Based on the results of our analysis, we discuss implications of our results and address our research questions.

\paragraph{Comparison to the Anonymization of Images and Regular Videos.}
In general, previous work indicates that participants value realistic or context-preserving anonymization like avatars, cartooning, or generative approaches and prefer them over blocking and blurring for images and videos~\cite{xu2024examining, li2017effectiveness, hasan2017cartooning, fernandes2023cartoonized}, unless a high level of privacy is required~\cite{mujirishvili2024don}.
Similarly, our results indicate that participants have negative opinions towards blocking.
It was rated as distracting and unpleasant with participants furthermore describing it as unnatural in the debriefing.
Another similarity to regular videos is that the detection accuracy for face-swapped 360$^\circ$ videos in both conditions is relative low with participants having problems to distinguish between original and altered videos~\cite{wohler:2020, wohler:2021}.

Regarding the perception of blurring and face-swapping, our results indicate some differences to regular videos.
In Screen condition, face-swapping is rated as most realistic and pleasant but not generally preferred over the other techniques.
Instead, our results suggest that blurring and face-swapping are similarly preferable.
These differences to previous studies are even more pronounced in HMD condition.
Here participants generally prefer blurring and we cannot detect differences between the pleasantness of blurring and face-swapping.
Even though participants described face-swapping as natural and difficult to recognize in the debriefing, they also discussed that face-swapping can introduce unpleasant and distracting artifacts.
As these more critical observations on face-swapping are increased in HMD condition and only seldom reported in portrait videos~\cite{woehler:2022}, it poses the question of whether face-swapping becomes more unpleasant due to the immersive properties of the HMD.
To further understand the effect of immersive viewing experiences, additional experiments focusing on a broader range of stimuli including comparisons between regular and 360$^\circ$ videos would be beneficial.

\paragraph*{Effectiveness of the Anonymization.}
As we neither test the ability of participants to actually identify people nor show direct comparisons of original and anonymized videos, we can only discuss the perceived effectiveness of facial anonymization in 360$^\circ$ videos.
Based on our data, blocking and blurring are visible to participants and overall make them feel as if the video is sufficiently anonymized.
In contrast, participants perceived face-swapping as less effective but this might not actually be the case as our results also highlight that people may not notice face-swapping.
Six participants stated that they cannot rate the efficiency of face-swapping well as they could not easily distinguish between non-anonymized and anonymized videos. 
This indicates that face-swapping may not be ideal in situations requiring transparent anonymization, e.g., for medical or criminal datasets.

Currently, there is no research on the actual anonymization effectiveness of facial anonymization in 360$^\circ$ videos.
While previous work on images suggest humans mostly cannot identify people for blocking and face-swapping, blurring seems less effective~\cite{li2017effectiveness, vishwamitra2017blur, khamis:2022}.
It would be valuable to confirm these results for 360$^\circ$ videos.

\paragraph*{Conspicuousness of Face-Swapping.}
Our results confirm the high quality and unobtrusiveness of face-swapping for 360° street videos.
Our stimuli show challenging real-world scenes, participants were aware of face-swapping, and the same target face is applied to all people, despite this the correct detection accuracy of participants was rather low with 32.5\% (Screen) and 42.5\% (HMD) which is comparable to portrait videos~\cite{wohler:2021}.
Additionally, some participants did not correctly identify any of the face-swapped videos (Screen 8, HMD 4 participants). 

Still, in the debriefing for the HMD condition participants mentioned unpleasant artifacts that can lead to discomfort. 
Mostly, participants noticed flickering when the face detection failed for single frames (4 participants) and faces that were not applied realistic (4 participants), see Fig.~\ref{fig:syntheticface} (right).
One participant reported mismatches depending on ethnic factors.
This could indicate that some face-body combinations introduce unappealing results which was briefly discussed for face-swaps between genders~\cite{wohler:2021}.
However, this work did not find face-swaps in portrait videos to generally reduce the appeal of the video.
As we especially find unpleasant artifacts in 360° face-swapped videos viewed in HMD the influence of immersive viewing might lead to a stronger perception of artifacts.
Overall, more investigations on artifacts and possible improvements of face-swapping should be considered in order to prevent discomfort.

\paragraph*{Balancing Realism and Privacy.}
One important aspect of video anonymization is the trade-off between privacy and video information~\cite{boyle2005language}.
The privacy required by users is generally dependent on the context of the recordings.
Previous work describes that user find blurring insufficient to protect their privacy for cameras installed in their living area or mounted to their body during the day~\cite{fernandes2023cartoonized, mujirishvili2024don}.
In our experiment, participants preferred blurring in HMD condition and found it similar effective as blocking indicating that they value a pleasant viewing experience and find the anonymization effect of blurring adequate for walking videos recorded in public areas.

In contrast to face-swapping, the widely used anonymization techniques blocking and blurring have a negative impact on the presence of viewers as the realism of the videos is reduced.
As increased presence is one of the main benefits of 360° videos, this effect could lead to a conflict of interest between the creators of immersive content and the necessity of anonymization as participants in previous studies on regular videos discussed that they would rather not use distracting anonymization~\cite{hasan2017cartooning}.
While face-swapping is the least obtrusive technique and does not generally reduce presence, participants still noticed artifacts which could introduce discomfort especially in HMD.
This indicates that further improvements on the technology might be necessary to deal with the challenging scenarios presented in 360° videos including variations in lighting and profile views of faces.

Additionally, as the realism of face-swapping increases, the technique might become fully undetectable by humans and machines.
While this could make the anonymization more efficient as viewers might not even question the identity of people~\cite{khamis:2022}, it might lead to challenges for videos that require a high level of privacy.
This especially needs to be addressed in videos of assisted living systems or medical scenarios which can benefit from preserving the original video context~\cite{wilson:2022, mujirishvili2024don}, but require verifiable anonymization.
Therefore, it is necessary to discuss how to verify that videos are anonymized and communicate the usage of anonymization to viewers.

\paragraph*{Ethical Consideration of Face-Swapping.}
Due to the highly realistic results and general low conspicuousness of face-swapping, the technique can be abused to create manipulative content by applying the facial appearance of non-consenting people to videos.
This has sparked many debates about the ethical use and dangers of the technique mostly focusing on defamatory or manipulative content of specific individuals.
It was discussed that face-swapped videos of politicians could be used to influence voting behavior and increase their mistrust~\cite{dobber:2020,zimmermann:2020}, or be used for various crimes by impersonating others~\cite{caldwell:2015}.
Furthermore, face-swapping has led to an increase in image-based sexual assault by applying faces of non-consenting people to pornographic imagery~\cite{flynn:2021} which can not only lead to a loss of reputation but even cause similar psychological harm as real-world abuse~\cite{mcglynn:2021}.
In the context of street videos, manipulations could target specific people by placing them at locations that they never actually visited to reduce their reputation, e.g., by applying one's face to pedestrians in a red-light district.

Moreover, street videos could be a target for manipulation by intentionally changing the facial appearance of bystanders to increase the appeal or change the impression of the video.
This way, creators might want to only use conventionally attractive people for face-swapping possibly introducing problems previously discussed for images in advertisement and social media which can negatively impact viewers' body image if only very attractive models~\cite{groesz:2002} or edited images~\cite{tiggemann:2020, kleemans:2018} are shown.
Furthermore, face-swapping can change the ethnic appearance of people in the video.
Therefore, it could be possible - either maliciously or due to a lack of consideration - to erase ethnic groups from videos.
In this case, the videos would not only fail to represent the diversity of the real world, but might also lead to negative impressions of minority groups which were observed for non-diverse virtual worlds~\cite{lee2011whose}.
Another factor unique for street videos is that creators cannot get feedback from recorded people on their preferences on the used anonymization.
While face-swapping can protect the privacy of bystanders, it might not actually be in their interest to have a different face applied to their body even in non-manipulative contexts.

Consequentially, it is important to consider ethical aspects even in context of the anonymization of street videos.
Moreover, due to their immersive properties and high realism, 360° videos could possibly increase negative effects of video manipulation which should be considered in future work.

\paragraph*{Demographic Factors.}
In our experiment, we recruit participants from different demographic groups for the Screen and HMD condition.
The participants in the Screen condition are recruited from the US and performed the experiment online, while the participants in the HMD condition are Japanese.
This can impact the comparability of both conditions.
In regards to presence, previous studies could not find significant differences between White and Asian participants, however, there were differences between Black and White people~\cite{martingano2023demographic}.
In contrast to our study, their participants were recruited from the same city.
As all of our stimuli show scenes of Japan, they may have a different effect based on whether participants live in that country or not especially considering the impression of familiarity. 
In addition, the higher detection accuracy of face-swapping in HMD condition might not necessary be related to the properties of the display but could be influenced by an own-race bias~\cite{chiroro1995investigation}.

\paragraph*{Limitations and Future Work.}
While our experiment provides valuable insights regarding facial anonymization of 360° videos, it also has some limitations.
First, as we performed the experiment in Screen condition online, we do not have full control over the final presentation of stimuli as the internet connectivity could reduce the resolution of the videos.
In cases where the video quality is significantly lowered, video anonymization using blurring and face-swapping may become less noticeable compared to the original 4K video files.
Finally, the experiment tells participants about the facial anonymization techniques beforehand and involves a detection task.
This means participants are aware of the anonymization and might focus more on artifacts than during natural viewing.
As this could influence their presence, further assessments focusing on natural video viewing would be beneficially.

\subsection{Discussion of Research Questions}

\paragraph*{RQ1: Which anonymization technique is least noticeable, distracting and most realistic?}
Similar to previous work on regular videos~\cite{wohler:2021, tahirdeepfake}, we find that distinguishing between face-swapping and original videos is difficult highlighting the realism of the technique for 360° videos.
Even though, we used the same facial identity for all face swaps, participants were seemingly not aware of this and only one participant commented that it was unnatural when several people had the same face in the debriefing.
Finally, blocking of faces with a solid box was perceived as most distracting and noticeable in line with previous work on photo and video anonymization~\cite{li2017effectiveness, fernandes2023cartoonized}.

\paragraph*{RQ2: Does facial anonymization impact the presence of participants?}
We only find differences between original and anonymized videos in HMD condition.
This may be due to higher presence when viewing 360° videos in HMD~\cite{filter2020virtual}, however, it is possible that other tasks or stimuli could elicit a difference in presence even in Screen condition which could be assessed in future experiments.
In the HMD condition, we find that especially blocking reduces the presence and scene impression and therefore might not be suited for immersive applications.

\paragraph*{RQ3: Which anonymization technique is best suited for 360° videos?}
Our results indicate that the best anonymization technique depends on the purpose of the videos.
Face-swapping is a promising technique as it does not impact the presence of users and retains the realism of the video.
It might be good for scenarios that require high realism, less distraction, or the preservation of facial expressions~\cite{wohler:2021,wilson:2022}.
However, it can also introduce artifacts causing discomfort to viewers.
Additionally, viewers are not easily able to recognize face-swapping making it difficult for them to verify whether the video is anonymized or maliciously altered.
Therefore, blurring could be preferable for videos that are challenging for face-swapping or require visible anonymization.
Finally, blocking is perceived mostly negative making it less useful for 360° videos.
\section{Conclusion}
Our results suggest that the perception of 360° videos is affected by anonymization.
Especially, for videos watched in HMD, facial anonymization can reduce the presence of viewers and negatively affect their impression of the scene.
We furthermore find significant differences in attitudes towards the tested anonymization techniques.
In general, face-swapping seems to be least noticeable and most realistic, hiding faces behind a black block was perceived as most negative, while blurring can offer a trade-off between realism and anonymization.
We additionally observe that the level of realism is not necessarily most important to viewers as they also consider how effective the facial identity is anonymized.
Consequently, the choice of anonymization technique needs to consider the video content and balance realism and verifiable privacy protection.
As our work only investigates the perceived anonymization effectiveness, it would be valuable to assess the actual anonymization effect for the techniques to ensure the privacy of recorded people.

\section*{Acknowledgments}
The authors gratefully acknowledge funding by the Japan Society for the Promotion of Science (JSPS KAKENHI 21H03460).

\bibliographystyle{unsrt}
\bibliography{References} 

\end{document}